\documentclass[12pt]{article}
\usepackage{txfonts}
\usepackage{amssymb}

\usepackage{scicite}

\usepackage{CJK}
\usepackage{subfigure}
\usepackage{times}
\usepackage{epsfig}
\usepackage{url}

\newenvironment{sciabstract}{%
\begin{quote} \bf}
{\end{quote}}

\newcounter{lastnote}

\title{  Dimensions, Structures and Security of Networks}

\author{ Angsheng Li$^{1}$, Wei Zhang$^{1,2}$, Yicheng Pan$^{1}$   \\
\normalsize{$^{1}$State Key Laboratory of Computer Science}\\
\normalsize{ Institute of Software, Chinese Academy of Sciences}\\
\normalsize{$^{2}$University of Chinese Academy of Sciences, P. R.
China} }

\date{}

\begin{document}


\baselineskip24pt


\maketitle

\begin{sciabstract}

One of the main issues in modern network science is the phenomenon
of cascading failures of a small number of attacks. Here we define
the dimension of a network to be the maximal number of functions or
features of nodes of the network. It was shown that there exist
linear networks which are provably secure, where a network is
linear, if it has dimension one, that the high dimensions of
networks are the mechanisms of overlapping communities, that
overlapping communities are obstacles for network security, and that
there exists an algorithm to reduce high dimensional networks to low
dimensional ones which simultaneously preserves all the network
properties and significantly amplifies security of networks. Our
results explore that dimension is a fundamental measure of networks,
that there exist linear networks which are provably secure, that
high dimensional networks are insecure, and that security of
networks can be amplified by reducing dimensions.

\end{sciabstract}

Network security has become a grand challenge in the current science
and technology. We proposed a new model of high dimensional networks
by natural mechanisms of homophyly, randomness and preferential
attachment. We found that low dimensional networks are much more
secure than that of the high dimensional networks, and that there
exists an algorithm to reduce dimensions of networks which preserves
network properties and significantly amplifies security of the
networks. Our model provides a foundation for both theoretical and
practical analyses of security of networks, and an approach to
amplifying security of networks.

Networks are proven universal topology of complex systems in nature,
society and industry~\cite{Ba2009}. One of the main issues of modern
network theory is that most networks are vulnerable to a small
number of attacks. This poses a fundamental issue of security of
networks~\cite{SFSRVR2009}.

  The first type of security is the connectivity security against physical attacks of removal of
  nodes. For this,
  it has been shown that in
scale-free networks of the preferential attachment (PA, for short)
model~\cite{Ba1999}, the overall network connectivity measured by
the sizes of the giant connected components and the diameters does
not change significantly under random removal of a small fraction of
nodes, but the overall connectivity of the networks are vulnerable
to removal of a small fraction of the high degree
nodes~\cite{AJB2000, CRB2000, M2004}.

The second type of security is the spreading security against
cascading failures by a small number of attacks. We notice that
cascading failures naturally occur in rumor spreading, disease
spreading, voting, and advertising etc~\cite{W2002,AM1991,M2000}. It
has been shown that in scale-free networks of the preferential
attachment model even weakly virulent virus can spread
\cite{PV2001}.

The authors have shown that cascading failures of attacks
are much more serious than that of the physical attacks of removal
of nodes, that neither randomness in the ER model \cite{ER1959,
ER1960} nor the preferential attachment scheme in the PA model
\cite{Ba1999} is a mechanism of security of networks, and that
homophyly and randomness together resist cascading failures of
networks ( A. Li, W. Zhang, Y. Pan and X. Li, Homophyly and randomness resist cascading failure in networks). This shows that some community structures play an
essential role in network security.

In practice, overlapping communities are omnipresent. This poses a
new question: What roles do the overlapping communities play in
network security?

In this article, we found that overlapping communities are obstacles
for security of networks. To solve this problem, we propose an
algorithm to amplify security of a network by reducing the dimension
of the network. The algorithm removes the obstacle of overlapping
communities in network security.

{\bf Results}

Let $G=(V,E)$ be a network. Suppose that each node $v\in V$ has a
threshold $\phi_v$. Let $S\subset V$ be a subset of vertices of $G$.
We define the infection set of $S$ in $G$
 recursively as follows: 1) initially we say that every node in $S$
 is {\it infected}, 2) a node $v\in V$ becomes infected, if $\phi_v
 $ fraction of $v$'s neighbors are already infected. We use ${\rm
 inf}^G(S)$ to denote the set of all nodes infected by $S$ in $G$.

{\bf Security Model}

We propose a new model of networks, the security model. It proceeds
as follows: Given a homophyly exponent $a$ and a natural number $d$,

\begin{enumerate}

\item [(1)] Let $G_d$ be an initial $d$-regular graph such that each node has a
distinct color and called seed.

For each step $i>d$, let $G_{i-1}$ be the graph constructed at the
end of step $i-1$, and $p_i=1/(\log i)^a$.

\item [(2)] At step $i$, we create a new node, $v$ say.

\item [(3)] With probability $p_i$, $v$ chooses a new color, in which case,

\begin{enumerate}

\item we call $v$ a seed,

\item (Preferential attachment) create an edge $(v,u)$ where $u$ is
chosen with probability proportional to the degrees of nodes in
$G_{i-1}$, and

\item (Randomness) create $d-1$ edges $(v,u_j)$, where each $u_j$ is
chosen randomly and uniformly among all seed nodes in $G_{i-1}$.

\end{enumerate}

\item [(4)] Otherwise, then $v$ chooses an old color, in which case,

\begin{enumerate}
\item (Randomness) $v$ chooses uniformly and randomly an old color as
its own color, and

\item (Homophyly and preferential attachment) create $d$ edges
$(v,u_j)$, where $u_j$ is chosen with probability proportional to
the degrees of all nodes of the same color as $v$ in $G_{i-1}$.
\end{enumerate}
\end{enumerate}

We denote the security model by $\mathcal{S}$. Let $G=(V,E)$ be a
network of model $\mathcal{S}$. Since every node $v\in V$ has only
one color, we define the dimension of $G$ to be $1$. In so doing, we
call $G$ a linear network.

It has been shown that: for sufficiently large
$n$, if $G$ is a network of the security model with $n$ nodes, then
almost surely (or with probability $1-o(1)$), the following
properties hold: (Proofs of the results are referred to A. Li, Y. Pan and W. Zhang, Provable security of networks).

\begin{enumerate}

\item [1)] (Power law) $G$ follows a power law.

\item [2)] (Small world property) The diameter of $G$ is
$O(\log n)$.

\item [3)] (Homophyly) Let $X$ be a homochromatic set of nodes. Then:

\begin{enumerate}

\item The induced subgraph $G_X$ of $X$ in $G$ is connected,

\item The diameter of $G_X$ is bounded by $(\log\log n)$,

\item $G_X$ follows a power law with the same power exponent as that
of $G$,

\item The degrees of nodes in $X$ follow a power law of the same
exponent as that of $G$,

\item The size of $X$ is $|X|=O(\log^{a+1}n)$, and

\item The conductance of $X$ in $G$ is $\Phi (X)=O(\frac{1}{|X|^{\beta}})$ for
some constant $\beta$, where $|X|$ is the size of $X$.

\end{enumerate}

\item [4)] (Uniform security) There exists an $\phi=o(1)$ such that for threshold $\phi_v=\phi$ for all
nodes $v$ in $G$, for any set $S$ of nodes of size within a
polynomial of $\log n$, the size of the infection set of $S$ in $G$
is $o(n)$.

\item [5)] (Random security) For every node $v$ in $G$, if $v$ defines its threshold randomly and
uniformly, i.e., $\phi_v=r/d_v$, where $d_v$ is the degree of $v$,
and $r$ is randomly and uniformly chosen from $\{1, 2, \cdots,
d_v\}$, then for any set $S$ of size bounded by a polynomial of
$\log n$, the size of the infection set of $S$ in $G$ is $o(n)$.

\end{enumerate}

1) - 3) demonstrate that networks of the security model have all the
useful properties of networks. 4) - 5) show that the networks of
model $\mathcal{S}$ are provably secure against cascading failures
of attacks.

The proofs of 4) and 5) involved new probabilistic and combinatorial
principles, for which we outline the ideas.

Let $G=(V,E)$ be a network constructed from $\mathcal{S}$. For a
node $v\in V$, we define the length of degrees of $v$ to be the
number of colors associated with all the neighbors of $v$, written
by $l(v)$. For a $j$, we define the $j$-th degree of $v$ to be the
size of the $j$-th largest homochromatic set among all the neighbors
of $v$. We use $d_j(v)$ to denote the $j$-th degree of $v$. Then
with probability $1-o(1)$, we have the following {\it degree
priority principle}:

(i) The length of degrees of $v$ is bounded by $O(\log n)$,

(ii) The first degree of $v$, $d_1(v)$ is the number of neighbors
that share the same color as $v$,

(iii) The second degree of $v$ is bounded by a constant $O(1)$, and

(iv) If $v$ is a seed node, then the first degree of $v$ is lower
bounded by, or at least, $\Omega (\log^{\frac{a+1}{4}}n)$.

A community is a homochromatic set $X$ of $V$, or the induced
subgraph of a homochromatic set $X$. We say that a community $X$ is
strong, if the seed node $x_0$ of $X$ cannot be infected even if all
its neighbors with colors different from that of $x_0$ are all
infected, unless there are non-seed nodes in $X$ which have already
been infected. Otherwise, we say that the community $X$ is
vulnerable. For appropriately chosen homophyly exponent $a$, the
properties (i) - (iv) above ensure that almost all communities of
$G$ are strong. Therefore, the number of vulnerable communities is
negligible.

Let us consider the infection among strong communities. Suppose that
$X$, $Y$ and $Z$ are strong communities with seeds $x_0$, $y_0$ and
$z_0$ respectively. It is possible that $x_0$ infects a non-seed
node $y_1\in Y$, $y_1$ infects the seed node $y_0\in Y$, and $y_0$
infects a non-seed node $z_1\in Z$. In this case, the infection of
the seed $x_0$ of $X$ generates a sequence of strong communities
$Y$, $Z$ and so on such that each of the communities contains
infected nodes. However by the construction of $G$, we have that
$x_0$ is created later than $y_1$, that $y_0$ is created later than
$z_1$, and that the edges $(x_0,y_1)$, $(y_0,z_1)$ are created by
the preferential attachment scheme at the time step at which a seed
node is created so that the edges are embedded in a tree $T$. We
call the tree $T$ the {\it infection priority tree} of $G$. The key
point is that the infection priority tree $T$ satisfies the
following basic principle: With probability $1-o(1)$, $T$ has height
bounded by $O(\log n)$. We call this property the {\it infection
priority tree principle}.

Therefore the infection of a seed node $x_0$ generates a path of at
most $O(\log n)$ many strong communities each of which contains
infected nodes. In addition, each community has size bounded by
$O(\log^{a+1}n)$. So even if all the nodes of an infected community
are infected, the contribution to the infection set is still
negligible.

For any attack $S$ of size bounded by a polynomial of $\log n$, let
$k$ be the number of vulnerable communities. Then there are at most
$|S|+k$ many seed nodes each of which generates a path of infected
strong communities. This allows us to prove that the size of the
infection set of $S$ in $G$ is $o(n)$, negligible comparing to $n$.

The arguments above show that the small community phenomenon of $G$
is one of the ingredients in the proofs of security of $G$. We
emphasize that the connecting patterns of the small communities are
crucial to the proofs of the security of $G$. In particular, the
degree priority principle ensures that almost all communities are
strong, and the infection priority tree principle ensures that any
path of infected strong communities has length $O(\log n)$.

Therefore linear networks from model $\mathcal{S}$ are provably
secure. In the proofs of the security result, the community
structures play an essential role in security of networks. However
it is not the case that every network with a community structure is
more secure.

{\bf $2$-Dimensional Security Model}

We generalize the security model to $2$ dimensions such that the
networks are rich in overlapping communities. The $2$-dimensional
security model proceeds as follows: Given a homophyly exponent $a$
and a natural number $d$,

\begin{enumerate}

\item [(1)] Let $G_d$ be an initial $d$-regular graph such that each node has a
distinct color and called seed.

For each $i>d$, let $G_{i-1}$ be the graph constructed at the end of
step $i-1$, and $p_i=1/(\log i)^a$.

\item [(2)] At step $i$, we create a new node, $v$ say.

\item [(3)] With probability $p_i$, $v$ chooses a new color, in which case,

\begin{enumerate}

\item we call $v$ a seed,

\item randomly and uniformly chooses an old color, $c$ say, as the
second color of $v$,

\item (preferential attachment) create an edge $(v,u)$ where $u$ is chosen
with probability proportional to the degrees of nodes in $G_{i-1}$,

\item for each $j=1,2,\cdots, d-1$:

- (randomness) with probability $\frac{1}{2}$, randomly and
uniformly chooses a seed node $u_j$, in which case, create an edge
$(v,u_j)$,

- (homophyly) otherwise, then chooses a node $u_j$ with probability
proportional to the degrees of nodes among all nodes sharing the
second color $c$ of $v$, and create an edge $(v,u_j)$.

\end{enumerate}

\item [(4)] Otherwise, then $v$ chooses an old color, in which case,

\begin{enumerate}
\item (randomness) $v$ chooses uniformly and randomly an old color as
its own color, and

\item (homophyly and preferential attachment) create $d$ edges
$(v,u_j)$, where $u_j$ is chosen with probability proportional to
the degrees of all nodes of the same color as $v$ in $G_{i-1}$.
\end{enumerate}

\end{enumerate}

We use $\mathcal{S}^2$ to denote the $2$-dimensional security model.
Let $G=(V,E)$ be a network constructed by $\mathcal{S}^2$. We define
a community of $G$ is the induced subgraph of a homochromatic set,
$X$ say. In this case, a seed node, $v$ say, may have two colors
$c_1$ and $c_2$, so that $v$ is contained in two communities $G_X$
and $G_Y$ say. Therefore, $G$ has an overlapping community
structure.

Here we interpret the maximal number of colors (or features) of
nodes for all the nodes of the network is the dimension of the
network. It is easy to extend the model to $k$-dimensional model for
arbitrarily given natural number $k$. We have argue that networks of
the security model $\mathcal{S}$ are secure. A natural question is:
Are networks of model $\mathcal{S}^2$ secure?

\begin{figure}[htbp]
                 \centering
                 \includegraphics[width=3.2in]{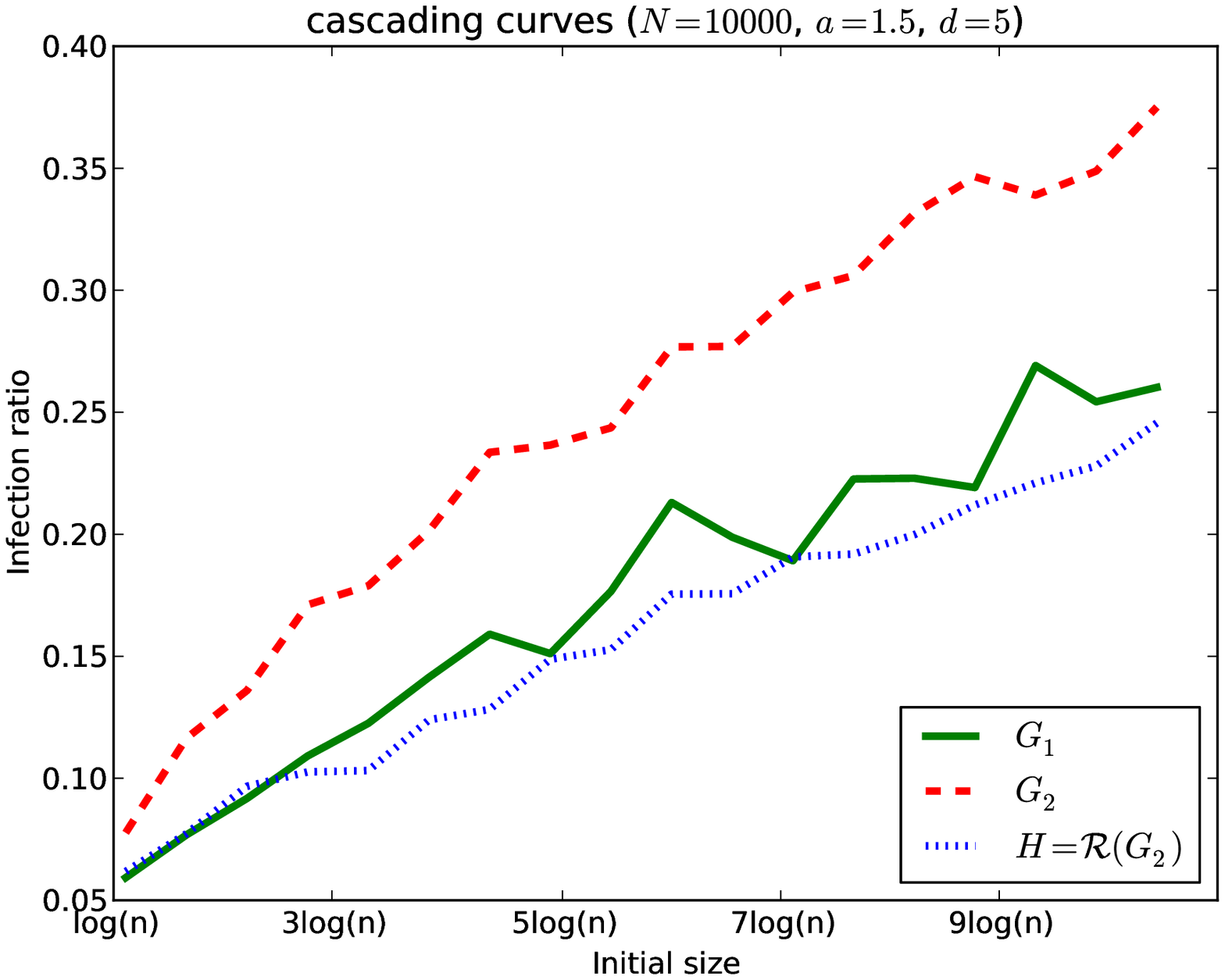}
                 \caption{Security curves. We use $G_1$ and $G_2$ to denote the networks
                  constructed from models $\mathcal{S}$ and $\mathcal{S}^2$ respectively. Let $H=\mathcal{R}(G_2)$. In the construction of
                  $G_1$ and $G_2$, $a=1.5$, $d=5$ and $n=10,000$. The horizon represents the number of attacked top degree nodes,
                  and the vertical line is the number of the largest infected sets among $100$ times of attacks. The security of $G_1$, $G_2$ and
                  $H=\mathcal{R}(G_2)$ are colored blue, green and red
                  respectively. In each time of attacks, the threshold of a node is randomly defined.}
                 \label{fig:split_overlapping_n=10000_d=5_a=1.5}
\end{figure}

\begin{figure}
                 \centering
                 \includegraphics[width=3.2in]{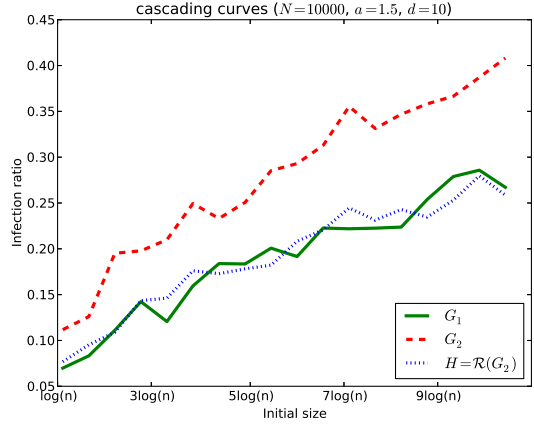}
                 \caption{Security curves. The same as Figure \ref{fig:split_overlapping_n=10000_d=5_a=1.5},
                  the difference is that in $G_1$ and $G_2$, $d=10$.}
                 \label{fig:split_overlapping_n=10000_d=10_a=1.5}

\end{figure}

In Figures \ref{fig:split_overlapping_n=10000_d=5_a=1.5} and
\ref{fig:split_overlapping_n=10000_d=10_a=1.5}, we depict the
security curves against a small number of attacks of the top degrees
of networks $G_1$'s from the security model $\mathcal{S}$, of
networks $G_2$ from the $2$-dimensional security model
$\mathcal{S}^2$. In both figures, $G_1$ and $G_2$ have homophyly
exponent $a=1.5$, and number of nodes $n=10,000$. The networks $G_1$
and $G_2$ in Figures \ref{fig:split_overlapping_n=10000_d=5_a=1.5}
and \ref{fig:split_overlapping_n=10000_d=10_a=1.5} have average
numbers of edges $d=5$, and $10$ respectively.

From Figures \ref{fig:split_overlapping_n=10000_d=5_a=1.5} and
\ref{fig:split_overlapping_n=10000_d=10_a=1.5}, we have that
networks of the security model are much more secure than that of the
$2$-dimensional security model. Therefore low dimensional networks
are more secure than that of the high dimensional networks. This
experiment shows that dimension is an important measure of networks
which plays an essential role in security of networks.

{\bf Algorithm $\mathcal{R}$: Reducing Dimensions of Networks}

We know that high dimensional networks are less secure than that of
low dimensional ones. Can we amplify security of networks by
reducing dimensions? For this, we introduce an algorithm to reduce
high dimensional networks to low dimensional ones.

Let $G=(V,E)$ be a network. Suppose that $\mathcal{X}=\{X_1,
X_2,\cdots, X_l\}$ is an overlapping community structure of $G$. We
introduce a graph reduction algorithm to remove the overlapping
communities of $G$ as follows.

\begin{enumerate}
\item [(1)] Let $X=\cup X_j$.

\item [(2)] For every $x\in X$, we split $x$ by the following steps:

\begin{enumerate}
\item suppose that $Y_1, Y_2,\cdots, Y_k$ are all communities $X_j$'s
containing $x$,

\item For each $i=1,2,\cdots,k$, let $d_i(x)$ be the number of
neighbors of $x$ that are in $Y_i$,

\item Replace $x$ by a circle of $k$ nodes $x_1,
x_2,\cdots, x_k$.

\item For each $i\in\{1,2,\cdots,k\}$, all the neighbors of $x$ that
are in $Y_i$ link to $x_i$.

\item For every neighbor $z$ of $x$ which is outside of $Y_i$ for
all $i$, with probability proportional to $d_i(x)$, we replace the
edge $(z,x)$ by $(z,x_i)$.

\end{enumerate}

\end{enumerate}

We use $\mathcal{R}$ to denote the reduction above. Clearly,
$\mathcal{R}$ splits the overlapping communities of $G$ into
disjoint communities.

{\bf $\mathcal{R}$ Preserves Network Properties}

\begin{figure}
  \centering

    \includegraphics[width=4in]{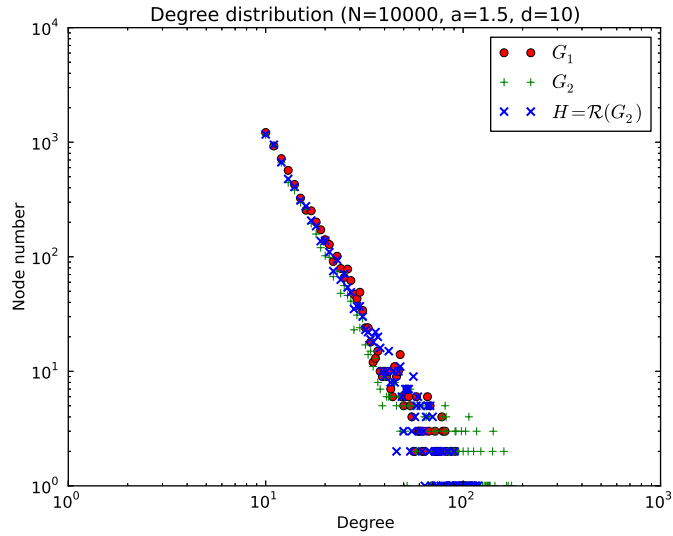}

\caption{Degree distributions of a network $G_1$ from $\mathcal{S}$
and a network $G_2$ from $\mathcal{S}^2$ and the reduced version of
$G_2$ by $\mathcal{R}$, that is, $H=\mathcal{R}(G_2)$, where the
homophyly exponent $a=1.5$, average number of edges $d=10$, and
number of nodes $n=10,000$ in the construction of $G_1$ and $G_2$.
The distributions of $G_1$, $G_2$ and $H$ are colored red, green and
blue respectively. From the figure, we know that $G_1$, $G_2$ and
$H$ all follow a power law of the same power
exponent.}\label{fig:degree_distri}

\end{figure}

In Figure \ref{fig:degree_distri}, we depict the degree
distributions of a network $G_1$ from the security model
$\mathcal{S}$, a network $G_2$ from the $2$-dimensional security
model $\mathcal{S}^2$ and the network $H$ reduced from $G_2$ by
$\mathcal{R}$, i.e., $H=\mathcal{R}(G_2)$, where the homophyly
exponent $a=1.5$, the average number of edges $d=10$ and the number
of nodes $n=10,000$ in the construction of $G_1$ and $G_2$. From the
figure, we know that $G_1$, $G_2$ and $H=\mathcal{R}(G)$ all follow
a power law of the same power exponent.

In Table \ref{tab:properties}, we report the diameters, average
distances and clustering coefficients of a network $G_1$ from the
security model $\mathcal{S}$, a network $G_2$ from the
$2$-dimensional security model $\mathcal{S}^2$, and the reduced
network $H=\mathcal{R}(G_2)$, where the homophyly exponent $a=1.5$,
average number of edges $d=10$, and $n=10,000$ for both $G_1$ and
$G_2$.

From Table \ref{tab:properties}, we have the following
properties:

\begin{enumerate}

\item The diameter and average number of distances of $G_1$ are approximately equal to that of $G_2$ respectively.

\item The diameter and average number of distances of
$H=\mathcal{R}(G_2)$ are slightly larger than that of $G_2$
respectively.

\item The clustering coefficient of $G_1$ is significantly larger
than that of $G_2$.

Therefore high dimensional networks have less clustering
coefficients than that of the low dimensional ones.

\item The clustering coefficient of $H=\mathcal{R}(G_2)$ is
approximately equal to that of $G_2$.

\end{enumerate}

We thus have that high dimensions significantly reduce the
clustering coefficients, and that high dimensions undermine security
of networks. However the high dimensions preserve both the power law
and the small world property of low dimensional networks. For the
reduction $\mathcal{R}$, we have:

\begin{enumerate}
 \item [1)] It reduces
dimensions of networks,

\item [2)] It preserves the clustering coefficients of networks,

\item [3)] It preserves the power law and small world property,  and

\item [4)] It slightly increases the diameters and average numbers of
distances of networks.
\end{enumerate}

4) is the only disadvantage of $\mathcal{R}$, for which we know that
the amount of increments of diameters and average numbers of
distances are small, compared to that of the origin networks.

{\bf $\mathcal{R}$ Amplifies Security of Networks}

In Figures \ref{fig:split_overlapping_n=10000_d=5_a=1.5} and
\ref{fig:split_overlapping_n=10000_d=10_a=1.5}, we depict the
security curves against a small number of attacks of the top degrees
of networks $G_1$'s from the security model $\mathcal{S}$, of
networks $G_2$ from the $2$-dimensional security model
$\mathcal{S}^2$, and the reduced networks  from $G_2$'s by
$\mathcal{R}$, i.e., the networks $H=\mathcal{R}(G_2)$. In both
figures, $G_1$ and $G_2$ have homophyly exponent $a=1.5$, and number
of nodes $n=10,000$. The networks $G_1$ and $G_2$ in Figures
\ref{fig:split_overlapping_n=10000_d=5_a=1.5} and
\ref{fig:split_overlapping_n=10000_d=10_a=1.5} have average numbers
of edges $d=5$, and $10$ respectively.

From Figures \ref{fig:split_overlapping_n=10000_d=5_a=1.5} and
\ref{fig:split_overlapping_n=10000_d=10_a=1.5}, we have the
following results:

\begin{enumerate}

\item For a network $G$ from model $\mathcal{S}^2$,
$H=\mathcal{R}(G) $ is much more secure than $G$.

\item For the same homophyly exponent $a$, average number of edges
$d$ and number of nodes $n$, let $G_1$ and $G_2$ be the networks
constructed from models $\mathcal{S}$ and $\mathcal{S}^2$
respectively. Then $H=\mathcal{R}(G_2)$ has approximately the same
security as that of $G_1$.

\end{enumerate}

These results demonstrate that algorithm $\mathcal{R}$ optimally
amplifies the security of networks of model $\mathcal{S}^2$ so that
it significantly amplifies security of networks against cascading
failures of attacks. We notice that the only cost of $\mathcal{R}$
is that it slightly increases the diameters and average numbers of
distances of networks.

{\bf Method Summary}

For the cascading failure of attacks, for a network $G=(V,E)$, every
node $v\in V$ picks randomly and uniformly a threshold
$\phi_v=r/d_v$, where $r$ is randomly and uniformly chosen from
$\{1,2,\cdots, d_v\}$, and $d_v$ is the degree of $v$ in $G$.
 The curves of cascading failures of all the experiments in Figures
\ref{fig:split_overlapping_n=10000_d=5_a=1.5} and
\ref{fig:split_overlapping_n=10000_d=10_a=1.5} are the maximal ones
among $100$ times of attacks of the top degree nodes.

{\bf Discussion}

Global cascading failure may occur in many networks by a small
number of attacks for which there are various reasons. Clearly the
$2$-dimensional model can be easily extended to higher dimensional
ones. However the $1$- and $2$-dimensional models are already
sufficient for us to investigate the roles of high dimensions in
structure and security of networks. We found that high dimensions
are the mechanisms of overlapping communities in networks, that high
dimensions and their structural characteristics of overlapping
communities are fundamental reasons of insecurity of networks
against cascading failures of attacks, and that there exists an
algorithm to amplify security of networks by reducing the
dimensions. Furthermore, our algorithm for reducing dimensions
preserve all the network properties of the original networks. The
new concepts and discoveries reported here include: 1) dimension is
a fundamental notion of networks, 2) there exist linear networks
which are provably secure, 3) high dimensions reduce clustering
coefficients and undermine security of networks, 4) security of
networks can be amplified by reducing dimensions, and 5) algorithms
for reducing dimensions preserve all the properties of the original
networks. These discoveries may not only point out new directions of
network theory, but also have implications and potentials in
understanding and analyzing complex systems in general.

\newpage

\begin{table}[ht]
\begin{center}
\begin{tabular}
{|c|c|c|c|} \hline
    & $G_1$ & $G_2$ & $H=\mathcal{R}(G_2)$ \\
\hline
Diameter & 10 & 9 & 12 \\
\hline
Average Distance & 5.68 & 5.19 & 6.33 \\
\hline
Clustering Coefficient & 0.535 & 0.359  & 0.352 \\
\hline
\end{tabular}
\end{center}
\caption{Network Properties. Diameters, average distances and
clustering coefficients of $G_1$, $G_2$ and $H=\mathcal{R}(G_2)$,
where $G_1$ and $G_2$ are constructed from the models $\mathcal{S}$
and $\mathcal{S}^2$ respectively, with $a=1.5$, $d=10$ and
$n=10,000$.} \label{tab:properties}
\end{table}

\newpage

{\bf Acknowledgements}

Angsheng Li is partially supported by the Hundred-Talent Program of
the Chinese Academy of Sciences. He gratefully acknowledges the
support of the Isaac Newton Institute for Mathematical Sciences,
Cambridge University, where he was a visiting fellow during the
preparation of this paper. All authors are partially supported by
the Grand Project ``Network Algorithms and Digital Information'' of
the Institute of Software, Chinese Academy of Sciences, and by an
NSFC grant No. 61161130530 and a
973 program grant No. 2014CB340302. Yicheng Pan is partially supported by a National Key Basic Research Project of China (2011CB302400)
and  the "Strategic Priority Research Program" of the Chinese Academy of
Sciences£¬Grant No. XDA06010701.

{\bf Author Contributions} AL designed the research and wrote the
paper, WZ, and YP performed the research. All authors reviewed the paper.

{\bf Additional information }

Competing financial interests: The authors declare they have no
competing financial interests.

\end{document}